\documentclass[12pt]{iopart}

\usepackage{subfigure}
\usepackage{graphicx}
\usepackage{dcolumn}
\usepackage{bm}
\usepackage{citesort}
\usepackage{multirow}

\def\ket#1{\left| #1 \right>}
\def\mel#1#2#3{\left< #1 \left| #2 \right| #3 \right>}
\def\kb#1#2{\left| #1 \right> \left< #2 \right|}
\def\bk#1#2{\left< #1 \left| #2 \right> \right.}

\begin{document}
\title{Control of atomic Rydberg states using guided electrons}

\author{T. Laycock, B. Olmos, T.W.A. Montgomery, W. Li, \\
T.M. Fromhold and I. Lesanovsky}
\address{School of Physics and Astronomy, The University of Nottingham, Nottingham,\\
 NG7 2RD, United Kingdom}

\date{\today}

\begin{abstract}
We present and analyze a simple model to illustrate the possibility of Rydberg state control by means of a moving guided electron. Specifically, we consider alkali metal atoms whose valence electron is initially prepared in a Rydberg $s$-state and investigate state changes induced by the interaction with the nearby passing electron. We analyze the dependence of the atomic final state, obtained after the passage, on the electron's momentum. We identify experimentally accessible parameter regimes and discuss the experimental feasibility of the proposed scheme. We furthermore discuss the possibility to excite and manipulate many-body states of a chain of interacting Rydberg atoms.
\end{abstract}


\maketitle

\section{Introduction}

Recent advances in the trapping and manipulation of ultracold atoms have opened up a plethora of new possibilities ranging from the study of fundamental quantum physics \cite{Bloch08} to the practical application of these controllable quantum many-body systems in quantum information processing and quantum simulation \cite{Brennen99,Brennen00,Jaksch00,Negretti11}. Currently there is growing interest in the use of atoms in highly excited Rydberg states \cite{Gallagher94,Lukin01,Mueller09,Saffman10,Pohl10,Weimer10,Schachenmayer10}. These atoms interact strongly through long-range dispersion forces \cite{Gallagher94} making them a promising platform for the study of many-body quantum phenomena and the implementation of quantum information processing protocols.

A central ingredient of any quantum computation or simulation scheme is the ability to control transitions between atomic states as this permits the initialization, addressing and characterization of single and many-body quantum states. For this purpose one usually employs coherent laser or microwave fields. The latter are particularly suited for the control of Rydberg atoms as transitions between nearby Rydberg states are typically in the microwave regime \cite{Raimond01,Haroche06,Guerlin07,Bohlouli07,Hogan12}.
\begin{figure}[ht]
\begin{center}
  \includegraphics[width=0.7\columnwidth]{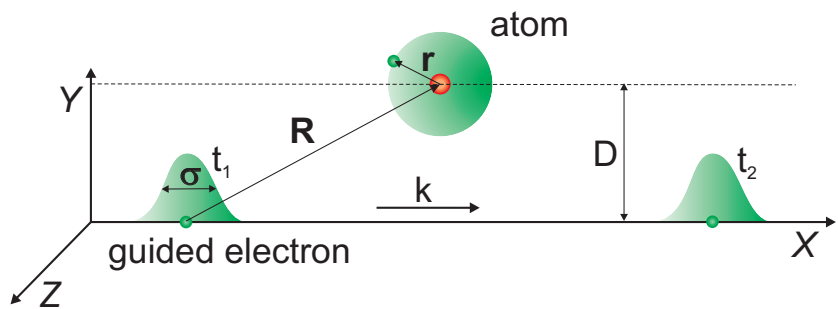}
  \caption{Schematic of the envisioned setup. A Rydberg atom is held at a distance $D$ from a quantum wire. Inside the wire an electron is propagating, which is modeled by a wave packet with central momentum $k$ and width $\sigma$. Initially (at time $t_1$), the electron is prepared far away from the atom such that they are non-interacting. We are interested in the state of the atom at time $t_2$ when the electron has passed near the atom and their interaction is again negligible.}\label{fig:scheme}
\end{center}
\end{figure}

In this work we discuss an alternative approach for Rydberg state control. The method we outline here relies on the interaction of the atom with the electric charge of a nearby guided electron (see Fig. \ref{fig:scheme}). This method links to current experimental and theoretical efforts that aim to explore the coupling of quantum devices to solid state systems \cite{Mozley05,Wallquist09,Petrosyan09,Judd10,Tauschinsky10,Sinuco11,Judd11,Cetina11,Wimberger11,Carter12,Hattermann12,Hogan12,Hogan12-1,Daniilidis13} and highlights the possibility to produce state changes of localized Rydberg atoms through electrons propagating in quantum wires or waveguides \cite{Hoffrogge11,HoffroggeNJP}. This approach can potentially find applications in the study and control of quantum many-body phenomena such as interaction-induced excitation transfer \cite{Mulken07,Wuster10,Wuster11} and also in quantum information processing protocols that rely on the switching of interacting Rydberg states.

The paper is structured as follows. In section \ref{System} we introduce a simple model describing a Rydberg atom interacting with a guided electron. In section \ref{Schrodinger} we detail how to obtain, under a number of approximations, the equation that describes the internal dynamics of the atom during the passage of the electron. By solving this equation numerically, we obtain the excitation probability of different atomic states after the passage of the electron for rubidium and lithium. We also provide an analytic approximation to this excitation probability valid in the limit of weak coupling between the atom and the electron and compare it with the numerical results. In section \ref{Many-Atoms} we outline the extension of the previous analysis to a chain of $N$ interacting Rydberg atoms and, finally, in section \ref{Experimental realization} we discuss the experimental implementation of the proposed scheme. We conclude with a summary and conclusions in section \ref{Summary}.

\section{System and Hamiltonian} \label{System}
The model we consider here is sketched in Fig. \ref{fig:scheme}. It consists of an electron confined to travel in the $X$ direction with momentum $k$ that approaches an atom trapped at a distance $D$ from the quantum wire. The Hamiltonian of this composite system is given by $H=H_\mathrm{E}+H_\mathrm{A}+H_\mathrm{int}$. The first term contains the kinetic energy of the electron (in atomic units) $H_\mathrm{E}=-\partial_X^2/2$, while the second one describes the internal state of the atom, $H_\mathrm{A}=\sum_{\alpha} E_{\alpha}\kb{\alpha}{\alpha}$, where $E_\alpha$ represents the energy of the atomic state $\ket{\alpha}$. We consider that the valence electron of the atom is initially in a highly excited orbit, moving in a central potential created by the positive core formed by the nucleus and the inner shell of electrons. Thus, the state of the atom can be labeled by its principal, orbital and azimuthal quantum numbers $n,l$ and $m$, respectively, $\ket{\alpha}\equiv\ket{nlm}$, with energies given by $E_\alpha=-1/2(n-\delta_l)^2$ where $\delta_l$ is the so-called quantum defect \cite{Gallagher94}. In our simple model description we neglect fine and hyperfine structure.

The last term of the Hamiltonian describes the Coulomb interaction between the components of the atom and the passing electron. We assume that $D\gg\left|\mathbf{r}\right|$, where $\mathbf{r}=(x,y,z)$ represents the relative coordinate of the valence electron. Hence, Taylor expanding accordingly the Coulomb potential of the involved charges, and separating the relative and center of mass coordinates, the interaction can be approximated as
\begin{equation*}
  H_\mathrm{int}\approx\frac{1}{2\left|\mathbf{R}\right|^3}\left[(x+iy)(X-iD)+\mathrm{h.c.}\right],
\end{equation*}
where $\mathbf{R}=(X,D,0)$ is the distance between the free electron and the atom. It is convenient to rewrite the interaction in terms of the atomic basis states as
\begin{equation}\label{eqn:Int_ham}
  H_\mathrm{int}=\frac{1}{2\left|\mathbf{R}\right|^3}\sum_{\alpha\alpha'}\left[\mu_{\alpha\alpha'}(X-iD)\kb{\alpha}{\alpha'}+\mathrm{h.c.}\right],
\end{equation}
where $\mu_{\alpha\alpha'} \equiv\mel{\alpha}{x+i y}{\alpha'}$ is the transition dipole matrix element between the atomic states $\ket{\alpha}$ and $\ket{\alpha'}$.

\section{The single-atom problem} \label{Schrodinger}
Initially, we consider that the electron is located very far away ($X\ll-D$) from the atom, which is prepared in a Rydberg $\ket{ns}$ state. As the electron approaches the atom, the interaction between the two systems induces a change in the internal state of the atom. We are interested in the excitation probability of each state after the interaction, when the electron is again very far from the atom ($X\gg D$). In this section, we solve the corresponding problem via the time-dependent Schr\"{o}dinger equation with Hamiltonian $H$ under a number of justified approximations.

The Schr\"{o}dinger equation reads
\begin{equation*}
  i\partial_t \psi(\mathbf{r},X,t)=H\psi(\mathbf{r},X,t),
\end{equation*}
where $\psi(\mathbf{r},X,t)$ represents the wavefunction of the system. The electron wavefunction is formed by a wave packet with central momentum $k$ (see Fig. \ref{fig:scheme}), i.e. $\psi(\mathbf{r},X,t) = f(\mathbf{r},X,t) e^{ikX}$. The Schr\"{o}dinger equation describing the dynamics of the coupled electron-atom system is
\begin{eqnarray}\label{eqn:envelope}
i \partial_t f(\mathbf{r},X,t)&=&-\frac{1}{2} \partial_X^2 f(\mathbf{r},X,t) -ik \partial_Xf(\mathbf{r},X,t)\\\nonumber
&&+ \left[V(\mathbf{r},X) +\frac{k^2}{2}\right]f(\mathbf{r},X,t)
\end{eqnarray}
where $V(\mathbf{r},X)=H_\mathrm{A}+H_\mathrm{int}$ includes the bare atomic energy levels and the atom-electron interaction. Assuming that the function $f$ varies slowly with $X$, i.e. $\left|\partial_Xf\right|\ll\left|kf\right|$ (slowly varying envelope approximation) we neglect the second order derivative in (\ref{eqn:envelope}). We may then rewrite it as
\begin{equation*}
  i \partial_t f(\mathbf{r},X,t)= -ik\partial_Xf(\mathbf{r},X,t)+V(\mathbf{r},X) f(\mathbf{r},X,t),
\end{equation*}
where the final term of (\ref{eqn:envelope}) has also been removed as it merely represents a global shift of the energies by $k^2/2$. This equation can be further simplified by a unitary transformation given by the operator $U=e^{-kt\partial_X}$ that brings us to a frame of reference where the electron is at rest. Hence, we reduce the problem to that of a particle at rest subject to a time dependent potential,
\begin{equation}\label{eqn:g}
i\partial_t g(\mathbf{r},X,t) =V(\mathbf{r},X+kt) g(\mathbf{r},X,t)
\end{equation}
with $g(\mathbf{r},X,t)=U^\dagger f(\mathbf{r},X,t)$. In the following we assume no back-action on the electron due to the state change of the atom, i.e. the change of the motional state of the electron due to the interaction with the atom is negligible. This is justified when the variance in momentum space of the electronic wavefunction is much greater than the smallest atomic energy level difference $\Delta_{n'l'}\equiv E_{n'l'}-E_{ns}$. As a consequence, we can separate the function $g(\mathbf{r},X,t)$ into $g(\mathbf{r},X,t)=\gamma(X)\phi(\mathbf{r},t)$, where $\gamma (X)$ and $\phi(\mathbf{r},t)\equiv\bk{\mathbf{r}}{\phi(t)}$ describe the electron and the atom, respectively. Here, the atomic state can be expanded as $\ket{\phi(t)}=\sum_{\alpha}C_\alpha(t)\ket{\alpha}$ with $C_\alpha$ representing the probability amplitude of each atomic state $\ket{\alpha}$.

Our aim is to obtain an effective equation of motion for the internal state of the atom. To do so, we multiply (\ref{eqn:g}) by $\gamma^*(X)$ and integrate over $X$ (the electron degree of freedom), so that we obtain
\begin{equation*}
i\partial_t \phi(\mathbf{r},t) =\int dX V(\mathbf{r},X+kt) \rho(X) \phi(\mathbf{r},t),
\end{equation*}
with $\rho(X)=|\gamma(X)|^2$. Finally, if the separation $D$ is much larger than the width of the electron envelope function $\sigma$, we can treat the electron as a point-like charge ($\rho(X)\approx \delta(X)$), and obtain
\begin{equation}\label{eqn:dynamics}
i\partial_t \phi(\mathbf{r},t) =V(\mathbf{r},kt)\phi(\mathbf{r},t).
\end{equation}
The solution of equation (\ref{eqn:dynamics}) will provide us the excitation probability of the different internal states of the atom after the passage of the electron. This solution can in general be obtained numerically, but one can also get an analytical approximation in the weak coupling limit, as we show in the following.

\subsection{Weak coupling regime} \label{sec:weak_coupling}

We consider in this section that the coupling between the atom and electron is so weak that during the passage of the electron the probability for the atom to undergo a state change is small. Since the initial state of the atom is $\ket{ns}$, and due to the selection rules of the dipole transitions ($l'=l\pm1$ and $m'=m\pm1$), the only accessible states in the weak coupling regime from the initial one are $\ket{n'p_{\pm}}\equiv\ket{n'p\,\pm 1}$. Hence, we can approximate the interaction Hamiltonian (\ref{eqn:Int_ham}) as
\begin{eqnarray*}
H_\mathrm{int}^{(1)}&=&\frac{1}{2\left|\mathbf{R}\right|^3}\sum_{n',n''}\mu_{n'n''} \left[(X-iD)\left(\kb{n's}{n''p_{-}}-\kb{n''p_{+}}{n's}\right)+\mathrm{h.c.}\right]
\end{eqnarray*}
where $\mu_{n'n''} \equiv \mu_{n's \,n''p_-}=-\mu_{n''p_+ \,n's}$ is the (positive and real) transition dipole moment between $\ket{n's}$ and $\ket{n''p_-}$. Inserting this Hamiltonian into Eq. (\ref{eqn:dynamics}) and expanding the atomic state as
\begin{equation*}
\ket{\phi(t)}=C_{ns}\ket{ns}+\sum_{n'}\left[ C_{n'p_+}\ket{n'p_+} + C_{n'p_-}\ket{n' p_-}\right] ,
\end{equation*}
one obtains the following set of coupled differential equations for the probability amplitudes of each state $C_{nl}$,
\begin{eqnarray*}
i\dot{C}_{n'p_+} & =& \lambda_{nn'}C_{n'p_+} - \eta_{nn'}{\cal F}(\tau) C_{ns} \\
i\dot{C}_{n'p_-} & =& \lambda_{nn'}C_{n'p_-} + \eta_{nn'}{\cal F}^*(\tau) C_{ns} \\
i\dot{C}_{ns} & = & \sum_{n'}\eta_{nn'}\left[{\cal F}(\tau)C_{n'p_-}-{\cal F}^*(\tau) C_{n'p_+}\right],
\end{eqnarray*}
with $\eta_{nn'}=\mu_{nn'}/(2D^2|\Delta_{n'p}|)$, $\lambda_{nn'}=\frac{\Delta_{n'p}}{|\Delta_{n'p}|}$ and ${\cal F}(\tau)=\frac{(\kappa\tau-i)}{\left[(\kappa\tau)^2+1\right]^{3/2}}$, written in terms of the dimensionless momentum $\kappa=k/(D|\Delta_{n'p}|)$ and time $\tau=t|\Delta_{n'p}|$.

The coefficient $\eta_{nn'}$ quantifies the coupling strength between the $\ket{ns}$ and $\ket{n'p}$ states. In the limit of very weak coupling ($\eta_{nn'} \ll 1$), it is justified to assume that the atom mainly remains in its initial state so that we put $C_{ns}\approx1$ and approximate the two first differential equations as
\begin{eqnarray*}
\dot{C}_{n'p_+} & =& -i\lambda_{nn'}C_{n'p_+} + i\eta_{nn'}{\cal F}(\tau) \\
\dot{C}_{n'p_-} & =& -i\lambda_{nn'}C_{n'p_-} - i\eta_{nn'}{\cal F}^*(\tau).
\end{eqnarray*}
We solve these equations with the initial condition $C_{n'p}(\tau=-\infty)=0$. The modulus squared of each amplitude when $\tau\to\infty$ represents the probability for the corresponding state to be populated after the passage of the electron. It yields
\begin{equation}\label{eqn:sigmap}
P_{n'p_\pm}=\left|C_{n'p_\pm}(\infty)\right|^2 =  4 \eta_{nn'}^2 \frac{1}{|\kappa|^4}\left[\lambda_{nn'}\frac{\kappa}{|\kappa|}K_0 \left(\frac{1}{|\kappa|}\right) \mp K_1 \left(\frac{1}{|\kappa|}\right) \right]^2
\end{equation}
where $K_n$ is the modified Bessel function of the second kind of order $n$. This expression is proportional to the square of the coupling strength parameter $\eta_{nn'}$, multiplied by a function that depends solely on the dimensionless momentum $\kappa$ and the sign of the energy difference between initial and final state $\lambda_{nn'}$. The function $P_{n'p_{\pm}}$ peaks at $\kappa^\mathrm{max} \approx \mp 0.7\times \lambda_{nn'}$ and the corresponding maximum of the transition probability is $\sim5\times\eta_{nn'}^2$.

\begin{figure}[ht]
\begin{center}
  \includegraphics[width=\columnwidth]{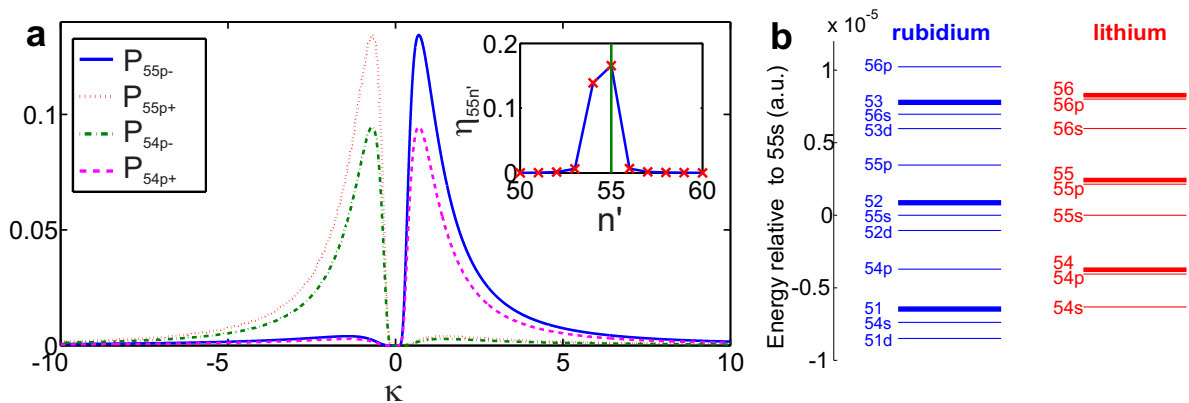}
  \caption{\textbf{a}: Transition probabilities from the initial $\ket{55s}$ state of a rubidium atom trapped at $D=2.5\,\mu$m from the electron guide as a function of its dimensionless momentum $\kappa$ given by Eq. (\ref{eqn:sigmap}). Represented are $P_{55p_+}$ (dotted red line), $P_{55p_-}$ (solid blue line), $P_{54p_+}$ (dashed purple line) and $P_{54p_-}$ (dash-dotted green line). The inset shows how the coupling of the various $\ket{n'p}$ states to the $\ket{55s}$ state ($\eta_{55n'}$) varies as a function of $n'$. \textbf{b}: Spectra of rubidium (blue) and lithium (red) around the $\ket{55s}$ state. The thick lines indicate the degenerate manifolds of high angular momentum states (states with zero quantum defect).}\label{fig:sigmann}
\end{center}
\end{figure}
Let us now discuss the transition probabilities in case of a rubidium atom initially prepared in the $\left|55s\right>$ Rydberg state. As is shown in the inset of Fig. \ref{fig:sigmann}a, the coupling strength $\eta_{nn'}$ is strongly dependent on the principal quantum number $n^\prime$ of the final state. Here $\eta_{nn}$ is the largest element, i.e., the probability is maximum for transitions with $n'=n$. A further dependence of the transition rates on the principal quantum number and the magnetic quantum number is introduced by the parameter $\lambda_{nn'}$ which is equal to $+1$ if $n'\geq n$ and $-1$ if $n'<n$. Hence, for $n'\geq n$, the $\ket{n'p_-}$ ($\ket{n'p_+}$) states are excited with maximum probability when $\kappa^\mathrm{max} \approx 0.7$ ($\kappa^\mathrm{max} \approx -0.7$) and vice versa for $n'<n$ (see, e.g., Fig. \ref{fig:sigmann}a). This shows that there is the possibility of experimentally selecting the internal magnetic quantum number of the excited atomic state solely by choosing the direction of the momentum of the electron.

Note, that the results of this section can also be obtained within a scattering theory framework, using the Lippmann-Schwinger equation \cite{Lippmann50}. The weak coupling limit is then equivalent to a treatment within the Born approximation, which describes only first order scattering events.

\subsection{Numerical results}

In order to move away from the weak coupling limit, we perform a numerical simulation of the dynamics given by Eq. (\ref{eqn:dynamics}). We consider two elements with very different spectra (see Fig. \ref{fig:sigmann}b), rubidium and lithium. Rubidium has large quantum defects whereas in the case of lithium only the $s$-state is appreciably separated from a quasi-degenerate manifold of states with higher angular momentum. We will analyze the similarities and differences in the state population of these two species after the passage of the electron.

\subsubsection{Analysis of excitation probabilities -} 
\begin{figure}[h]
\centering
\includegraphics[width=\columnwidth]{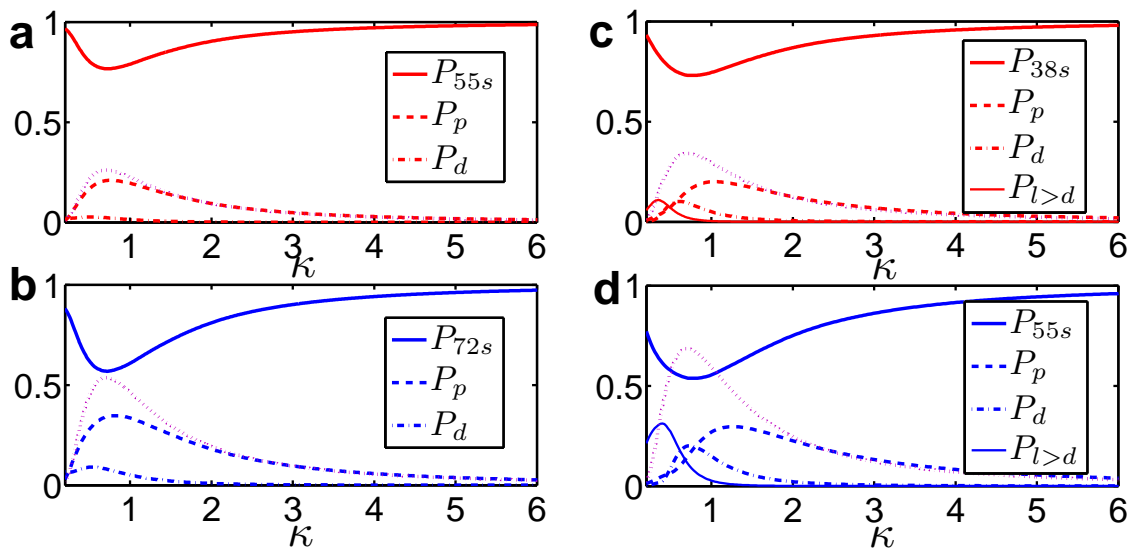}
\caption{Numerically calculated initial state population $P_{ns}$ and transition probabilities $P_p, P_d$ and $P_{l>d}$. The dotted lines represent the transition probability to a $p$-state obtained in section \ref{sec:weak_coupling} in the weak coupling limit as a function of the scaled momentum $\kappa$. \textbf{a}: Rubidium initial state $\left|55s \right>$ ($\eta_{55\,55}=0.18$), \textbf{b}: Rubidium initial state $\left|72s \right>$ ($\eta_{72\,72}=0.25$), \textbf{c}: Lithium initial state $\left|38s \right>$ ($\eta_{38\,38}=0.25$) and \textbf{d}: Lithium initial state $\left|55s \right>$ ($\eta_{55\,55}=0.35$).}
\label{Fig:RuLi}
\end{figure}

In Fig. \ref{Fig:RuLi}, we show the overall probability for an atom to remain in the initial state ($P_{ns}$) and the probability of occupation of states with a defined orbital quantum number $P_l=\sum_{n'm}P_{n'lm}$ for rubidium and lithium and two different values of the principal quantum number of the initial state.

Let us first comment on the features both elements have in common. The probability for the atom to remain in the initial state is lower the larger the coupling strength, $\eta_{nn'}$, which is an increasing function of the principal quantum number $n$. It shows a minimum for intermediate values of the scaled momentum $\kappa\sim1$. The shape of the transition probabilities to the $p$- and $d$-states ($P_p$ and $P_d$, respectively) are qualitatively very similar. However, the maximum value of $P_d$ is systematically smaller and sits at a lower value of $\kappa$. This is expected, as the atom would have to perform at least two transitions in order to end up in these states, and a lower momentum of the electron would favor these second order (slower) processes due to a longer interaction time with the atom. Even lower values of the momentum are required to find the atom in a state with higher orbital angular momentum ($P_{l>d}$).

In Fig. \ref{Fig:RuLi} we also compare the numerical value of $P_p$ with the analytic results obtained in the weak coupling regime in Sec. \ref{sec:weak_coupling} (dotted line in all panels). One can observe that, as expected, the lower the value of the coupling strength the better the agreement between the two curves, since the analytic result is only valid when $\eta_{nn'}\ll1$. However, the analytic prediction that reproduces qualitatively the main features of the rubidium case does not do so in lithium, where the maxima of the numerical transition probabilities are placed systematically at larger values of $\kappa$. The reason for this discrepancy is to be found in the differences between the spectra of the two species (Fig. \ref{fig:sigmann}b). To obtain the analytic results in Sec. \ref{sec:weak_coupling}, we make the approximation of only considering the coupling between the initial $s$- and the $p$-states. Due to the small quantum defects in lithium, this approximation breaks down due to the small energetic gap between the $p$- and the higher $l$-states, that makes higher order transitions more likely than in rubidium.

\subsubsection{Creation of permanent electric dipoles -} Due to its symmetry the initial Rydberg $s$-state does not carry a permanent electric dipole moment. However, the change of the population of the Rydberg states under the action of the passing electron induces such permanent dipole moment or polarization.

\begin{figure}[h]
\centering
\includegraphics[width=\columnwidth]{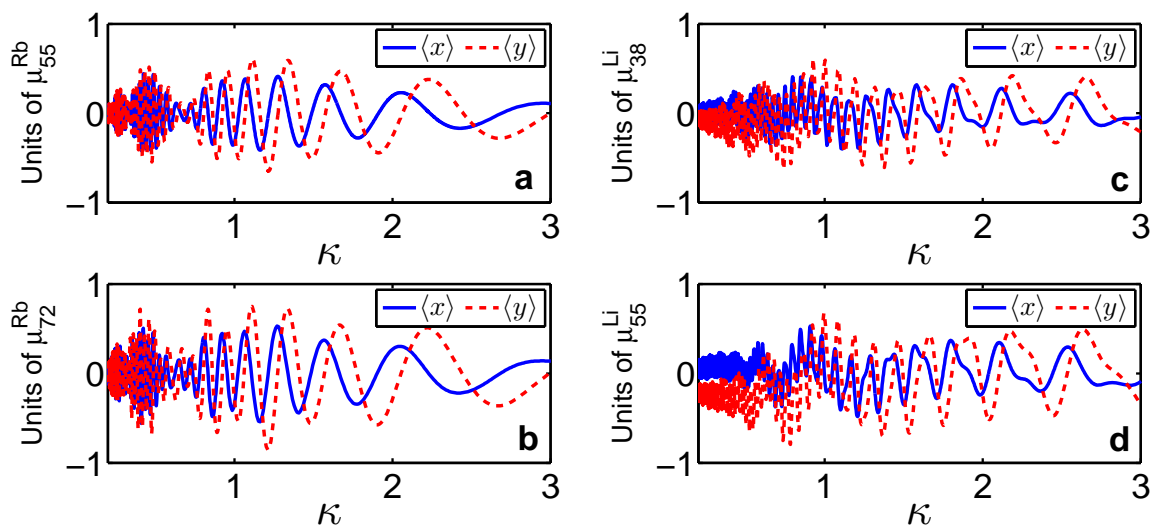}
\caption{Polarization of the final atomic state after the interaction for \textbf{a}: Rubidium initial state $\left| 55s \right>$ \textbf{b} Rubidium initial state $\left| 72s \right>$ \textbf{c}: Lithium initial state $\left| 38s \right>$ and \textbf{d}: Lithium initial state $\left| 55s \right>$. The parameters are the same as in Fig. \ref{Fig:RuLi}.}
\label{Fig:Polarization}
\end{figure}

The polarization in each direction is quantified by the expectation values $\left< x \right>$, $\left<y\right>$ and $\left< z \right>$, respectively. Shown in Fig. \ref{Fig:Polarization} is the polarization in both $x$ and $y$ directions for the values of the coupling strength used before in Fig. \ref{Fig:RuLi}. There is zero polarization in the $z$-direction ($\left< z \right>=0$) as the electron-atom interaction Hamiltonian has no component in this direction due to the geometry of the system.

Let us first comment on the features observed in the case of rubidium (Figs. \ref{Fig:Polarization}a and b). For $\kappa\leq 0.5$, both $\left<x\right>$ and $\left<y\right>$ oscillate very rapidly around zero, with maximum amplitude approximately equal to $0.5\times\mu^\mathrm{Rb}_{n}$. With increasing $\kappa$, however, the polarization becomes slightly elliptical, as the amplitude of $\left<y\right>$ becomes larger than that of $\left<x\right>$. We also observe that for $\kappa > 2$ the polarization decreases towards zero, as expected since here transitions are highly suppressed, i.e. $P_{55s}\approx 1$. For the values of the coupling strength represented here, one is still within the weak coupling regime where the perturbative analysis is valid. As a consequence, the qualitative features of $\left<x\right>$ and $\left<y\right>$ seem to be very similar. However, increasing the principal quantum number does enhance the polarization of the atom, consistent with a higher transition probability from the initial state. In lithium, however, the results are very different from the rubidium ones (see Figs. \ref{Fig:Polarization}c and d). While the mean value of the polarization decreases to zero for high $\kappa$ as was the case in rubidium, for low $\kappa$ it is positive for $\left<x\right>$ and negative for $\left<y\right>$. The separation between the two mean values gets larger with increasing principal quantum number $n$, indicating a larger value of the final polarization. This is consistent with the fact that the excitation probability $P_{l>d}$ to a high orbital momentum state is quite large for low values of $\kappa$ in Li (see Figs. \ref{Fig:RuLi} c and d).

This analysis shows that a passing electron can 'switch on' a permanent dipole moment in the atom which is of the order of the transition dipole between neighboring Rydberg states and thus can reach several thousand Debye. This control on the single-atom level can be used, for example, to explore the interaction-induced transfer of a single excitation in a many-body system \cite{Mulken07,Wuster10,Wuster11}.

\section{Extension to a chain of many atoms} \label{Many-Atoms}
\begin{figure}[h]
\centering
\includegraphics[width=\columnwidth]{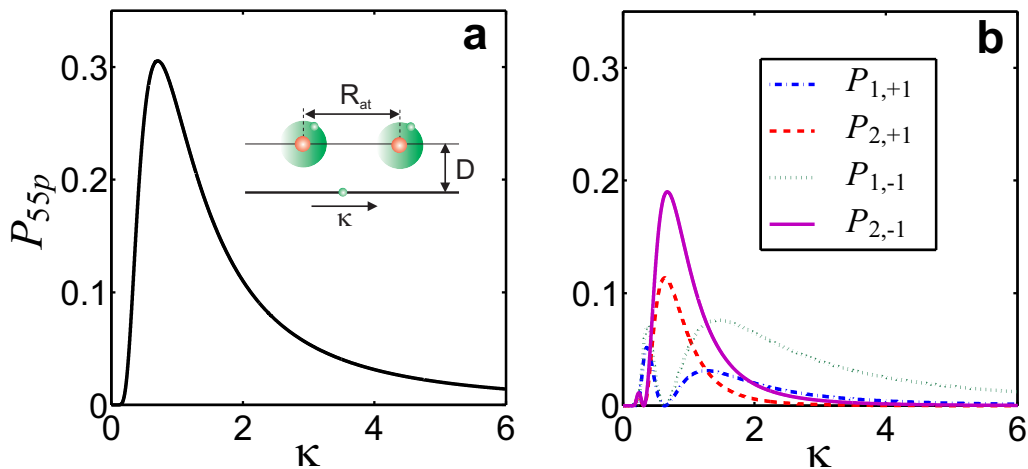}
\caption{Transition probabilities for two Rb atoms with $D = 2.5 \mu$m and $R_\mathrm{at} = 2D$ from the initial state $\left| 55s, 55s \right>$ to a many-body state with a single $p$-state excitation and $n'=n=55$. \textbf{a}: Overall probability for the atom changing its state. \textbf{b}: Probability of excitation of each eigenstate $\ket{m,\chi}$.}
\label{Fig:all_trans}
\end{figure}

We finally outline how to extend our analysis to a chain of $N$ interacting Rydberg atoms that are positioned parallel to the wire and separated from each other by a distance $R_\mathrm{at}$ (see inset of Fig. \ref{Fig:all_trans}a).

The interatomic interaction is given by the dipole-dipole potential
\begin{equation*}
V_\mathrm{dd}=\frac{1}{R_\mathrm{at}^3}\sum_{i\neq j}^N\frac{1}{|i-j|^3}\left(\mathbf{r}_i\cdot \mathbf{r}_j - 3 x_ix_j\right),
\end{equation*}
where $\mathbf{r}_i=(x_i,y_i,z_i)$ with $i=1\dots N$ is the relative coordinate of the valence electron of the $i$-th atom. In order to gain some general insights into the system we make a number of approximations to simplify the problem. Due to the $1/R^3$ dependence of $V_\mathrm{dd}$, we assume that the interaction between atoms separated further than nearest neighbors is negligible. Furthermore, we consider only resonant terms of the dipole-dipole interaction and restrict ourselves to the weak coupling regime, i.e. at most only a single atom can be excited to a $p$-state. Within these approximations, and abbreviating $\ket{n'p_\pm}_i\equiv \ket{ns}_1\otimes...\otimes\ket{n'p_\pm}_i\otimes...\otimes\ket{ns}_N$, the operator of the dipole-dipole interaction is given by
\begin{eqnarray}\label{eqn:dipole}
V_\mathrm{dd}^{(1)} = -\frac{1}{4R_\mathrm{at}^3} \sum_{n'n''}&&\mu_{nn'}\mu_{nn''} \sum_{i=1}^N \left[\left|n'p_+\right>_{i\,i+1}\!\!\left<n''p_+\right|+\left|n'p_-\right>_{i\,i+1}\!\!\left<n''p_-\right|\right.\\\nonumber &&\left.-3\left|n'p_+\right>_{i\,i+1}\!\!\left<n''p_-\right|-3\left|n'p_-\right>_{i\,i+1}\!\!\left<n''p_+\right| +\mathrm{h.c.} \right].
\end{eqnarray}
We are now interested in the collective states in which $V_\mathrm{dd}^{(1)}$ is diagonal as these are the 'good' eigenstates of the atomic system before and after the passage of the electron. They are given by the exciton states
\begin{equation*}
\ket{m,\chi}=\sqrt{\frac{2}{N+1}}\sum_{j=1}^N \sin \left[\frac{mj\pi}{N+1}\right] \frac{\ket{n'p_+}_j+\chi\,\ket{n'p_-}_j}{\sqrt{2}},\qquad\mathrm{for}\quad m=1\dots N,
\end{equation*}
with $\chi=\pm 1$ and the corresponding eigenvalues read
\begin{equation*}
\mathcal{E}_{m \chi} = \Delta_{n'p} - (1-3\chi) \frac{\mu_{nn'}^2}{2R_\mathrm{at}^3}\cos\left[\frac{m\pi}{N+1}\right].
\end{equation*}

We are interested in the transition probabilities $P_{m,\chi}$ from the initial state $\ket{ns}_1\otimes...\otimes\ket{ns}_N$ to each of the collective eigenstates $\ket{m,\chi}$. These can, like in the single-atom case, be calculated analytically under a number of assumptions. First, we assume that the electron is initially traveling in the positive $X$-direction ($\kappa>0$). We furthermore assume that the kinetic energy of the electron is much larger than $\Delta_{n'p}$ and that $\mu_{nn'}^2/R_\mathrm{at}^3\ll \Delta_{n'p}$ which is anyway required for Eq. (\ref{eqn:dipole}) to be valid. Within these approximations the transition probability yields
\begin{eqnarray}\label{eqn:multi}
P_{m,\chi} &\approx & \frac{4\eta_{nn'}^2}{N+1} \frac{1}{\kappa^4}\left[ K_1^2\left(\frac{1}{\kappa}\right) \delta_{\chi,{-1}} + K_0^2\left(\frac{1}{\kappa}\right) \delta_{\chi,{+1}} \right]\\\nonumber
&&\times\left[ \frac{\sin\left(\alpha_m\frac{N}{2}\right)}{\sin\left(\frac{\alpha_m}{2}\right)}  +(-1)^{m+1}\frac{\sin\left(\beta_m\frac{N}{2}\right)}{\sin\left(\frac{\beta_m}{2}\right)}\right]^2,
\end{eqnarray}
where
\begin{equation*}
\alpha_m=\frac{m\pi}{N+1} + \lambda_{nn'}\frac{R_\mathrm{at}}{\kappa D}
\quad
\mathrm{and}
\quad
\beta_m=\frac{m\pi}{N+1} - \lambda_{nn'}\frac{R_\mathrm{at}}{\kappa D}.
\end{equation*}
Note that the structure of the excitation probability  (\ref{eqn:multi}) is quite similar to the expression that was obtained in the single atom case (\ref{eqn:sigmap}). However, the many-body expression contains an extra oscillating factor which depends on both, the momentum $\kappa$ and the ratio $R_\mathrm{at}/D$ between the atomic separation and the distance to the electronic guide.

The total probability for the atom changing its state is given by summing over all final states
\begin{eqnarray*}
P_{n'p}=\sum_{m,\chi}P_{m,\chi} &\approx & 8N\eta_{nn'}^2\frac{1}{\kappa^4}\left[ K_1^2\left(\frac{1}{\kappa}\right) + K_0^2\left(\frac{1}{\kappa}\right)\right].
\end{eqnarray*}
This is simply $N$ times the total single atom transition probability $P_{n'p_+}+P_{n'p_-}$ [see Eq. (\ref{eqn:sigmap})] of the single atom case.

In Fig. \ref{Fig:all_trans} we provide the data calculated from (\ref{eqn:multi}) for $N=2$ atoms initialized in the $55s$-state. In panel (a) we show the total transition probability and in panel (b) the transition probability to each of the four many-body eigenstates of $V_\mathrm{dd}^{(1)}$ as a function of the electron's momentum. Here we can observe that the maximum transition probability corresponds to the antisymmetric state $\ket{2,-1}$. The data moreover shows that at the points where the probabilities for the transition to $\ket{2,-1}$ and $\ket{2,+1}$ peak, the excitation probabilities to the remaining states are zero. This shows that, at least to some extent, some selectivity of the excitation is possible by controlling the momentum of the electron.

\section{Experimental implementation} \label{Experimental realization}

We now consider possible routes to studying the effects that we predict in experiment. The excitation of atoms to Rydberg states in optical lattices \cite{Raithel11,Viteau11,Schauss12} and dipole traps \cite{Gaetan09,Urban09} has been demonstrated recently in a number of experiments. Hence, the key requirement for our scheme is the availability and control of electrons confined to move in one direction with a well defined momentum. In this respect, the recent development of chip-based systems, in which low-energy beams of free electrons are guided above the chip surface \cite{Hoffrogge11,HoffroggeNJP}, are particularly promising. A major advantage of this type of device is that the electrons are not embedded within the chip and, hence, their coupling to adjacent Rydberg atoms would not be significantly influenced either by the chip material (e.g. Casimir-Polder atom-surface attraction) or by other conduction electrons.

The typical values for the kinetic energy of single electrons moving in these kinds of waveguides are on the order of a few eV. As shown in Table \ref{tab:kin}, in this kinetic energy range our proposed scheme yields a transition probability from the initial state between $10\%$ and $20\%$. In all cases considered the transition to a $p$-state is the most probable, with a probability that ranges from $10\%$ to $18\%$.
\begin{table}[h]
\centering
\begin{tabular}{cc|c|c|l}
\cline{2-4}
& \multicolumn{1}{ |c| }{$n$} & $E_\mathrm{kin} (10\%)$ & $E_\mathrm{kin} (20\%)$ & \\ \cline{1-4}
\multicolumn{1}{ |c }{\multirow{2}{*}{Rb} } &
\multicolumn{1}{ |c| }{55} & 1.28 & 0.39 &  \\ \cline{2-4}
\multicolumn{1}{ |c  }{} &
\multicolumn{1}{ |c| }{72} & 1.55 & 0.65 &  \\ \cline{1-4}
\multicolumn{1}{ |c }{\multirow{2}{*}{Li} } &
\multicolumn{1}{ |c| }{38} & 1.74 & 0.74 &  \\ \cline{2-4}
\multicolumn{1}{ |c  }{} &
\multicolumn{1}{ |c| }{55} & 1.56 & 0.54 &  \\ \cline{1-4}
\end{tabular}
\caption{Table showing the electron kinetic energies (eV) required for the probabilities of a transition from the initial state to be $10\%$ and $20\%$.}\label{tab:kin}
\end{table}

However, free electron beams of this sort are currently not widely available. By contrast, electrons moving in one-dimensional quantum wires fabricated in two-dimensional electron gases (2DEGs) within semiconductor heterostructures or graphene are now routinely available in many laboratories worldwide \cite{Ando82,Castro09,Eva12}. Due to the major advances in materials and device fabrication techniques that have occurred over the last 30 years, the quality of such structures, as measured by the mobility and quantum behavior of the electrons, is extremely high. Moreover, by using etching or negative bias voltages applied to surface gates locally to deplete electrons, quantum wires can be made sufficiently narrow to support a single quantized conductance channel with well-defined spatial width and electron propagation velocity. Recent work has highlighted the potential for using conduction channels within 2DEGs to trap and/or detect ground-state atoms \cite{Judd10,Sinuco11}. Due to their high polarizability, even single Rydberg atoms should interact strongly with electrons in this type of quantum electronic structure.

In GaAs/(AlGa)As heterojunctions, the 2DEG is typically located 40-100 nm below the surface. Even the thin dielectric layer between the electrons and the surface significantly weakens the coupling of electrons to surface potentials \cite{Judd10}. However, in InAs-based samples the 2DEG forms on the surface itself \cite{Ando82}. Quantum wires fabricated within the 2DEG would then create one-dimensional transport channels in which electrons could propagate and directly couple to nearby Rydberg atoms without being separated by other material or electrons. However, further studies are needed to determine whether the analysis presented in this work is valid for electrons in these structures.

In the longer term, free-standing graphene membranes \cite{Du09}, in which it may be possible to create and isolate conduction channels by separating them with insulating graphane tracks \cite{Wei10}, could be an ideal system for studying the coupling of Rydberg atoms to electrons. For example, the Casimir-Polder attraction is expected to be low and there is great control over the charge carriers, which can be made electron or hole-like by applying an electric field normal to the layers \cite{Castro09,Eva12,Bordag06,Bordag09}. Such structures look promising for electronic imaging of deposited atom clouds \cite{Judd11} and may be able to detect Rydberg atoms above the surface. In this type of structure, the energy-wavevector dispersion relation of the charge carriers can also be changed from linear (Dirac fermion) to parabolic (free electron-like) by changing the width of the conducting channel or the number of graphene sheets. This control may yield additional advantages for tuning and measuring the interaction of mobile charges with Rydberg atoms.

\section{Summary and conclusion} \label{Summary}
We have studied how a guided electron may be used to manipulate the internal state of a Rydberg atom. After establishing a simple model description and using certain approximations we have derived an equation describing the time evolution of the atomic wavefunction under the action of a passing electron. This equation was solved analytically in the weak coupling limit, whose results predict that the direction from which the incident electron impinges on the atom selects which magnetic sublevel is populated. We have moreover conducted a numerical analysis of this equation and compared the results for the two elements lithium and rubidium. Beyond transition probabilities we have calculated the polarization of an atom after the electron has passed by. We have shown that a polarization on the order of the transition dipole moment between neighboring Rydberg $s$ and $p$-states is achievable. Finally, we have considered a system composed of $N$ interacting atoms that are aligned parallel to the electron guide. Here we showed that by tuning the electron's momentum, some selectivity in the excitation of collective many-body states can be achieved.

The results in this work have been obtained under a number of rather crude approximations. They, however, give a first indication that the control of Rydberg states through guided electrons might be a useful tool for the manipulation of, and also for probing, quantum many-body states. In the future it will be interesting to investigate whether this method can find applications for instance in hybrid quantum information processing architectures.

\ack
  The authors acknowledge funding by EPSRC. BO acknowledges funding by the University of Nottingham. IL and WL acknowledge funding from the ERA-NET CHIST-ERA (R-ION consortium).
\\
\providecommand{\newblock}{}

\end{document}